\documentclass[conference]{IEEEtran}
\IEEEoverridecommandlockouts
\usepackage{cite}
\usepackage{amsmath,amssymb,amsfonts}
\usepackage{algorithmic}
\usepackage{graphicx}
\usepackage{textcomp}
\usepackage{xcolor}
\usepackage{multirow}
\usepackage{amsmath}
\usepackage{latexsym}
\usepackage{caption}
\usepackage{hyperref}

\def\BibTeX{{\rm B\kern-.05em{\sc i\kern-.025em b}\kern-.08em
    T\kern-.1667em\lower.7ex\hbox{E}\kern-.125emX}}
\begin{document}

\title{Read, Extract, Classify: A Tool for Smarter Requirements Engineering
}

\author{\IEEEauthorblockN{\textbf{Paheli Bhattacharya, Manojit Chakraborty, Santhosh Kumar Arumugam, Rishabh Gupta}}
 \IEEEauthorblockA{\textit{Bosch Research and Technology Centre, Bangalore, India}}
 [paheli.bhattacharya, manojit.chakraborty, santhoshkumar.arumugam, gupta.rishabh]@in.bosch.com}

\maketitle

\begin{abstract}
This paper presents the ReXCL tool, which automates the extraction and classification processes in requirements engineering, enhancing the software development life-cycle. The tool features two main modules: Extraction, which processes raw requirement documents into a predefined schema using heuristics and predictive modeling, and Classification, which assigns class labels to requirements using adaptive fine-tuning of encoder-based models. The final output can be exported to external requirement engineering tools. Performance evaluations indicate that ReXCL significantly improves efficiency and accuracy in managing requirements, marking a novel approach to automating the schematization of semi-structured requirement documents. 
\end{abstract}


\begin{figure*}[t]
\caption{ReXCL pipeline. Input is customer requirement. Extraction module parses the document to produce structured tabular output. Classification module classifies each requirement text. The final output can be exported to the tools like IBM Doors.}
\label{fig:overall_workflow}
\centering
\includegraphics[width = \textwidth]{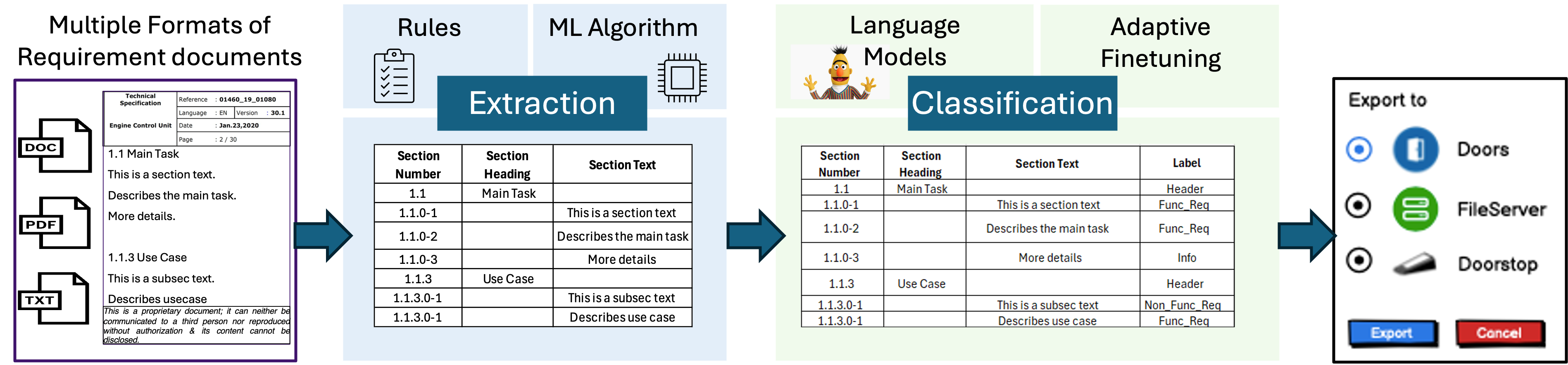}
\end{figure*}



\section{Introduction}
In software development, effective requirements engineering (RE) is essential to ensure that a system meets the expectations of diverse stakeholders --- including customers, developers, project managers, and systems engineers~\cite{ser1}. These stakeholders rely on structured requirement documents to guide project planning, development, and validation. Typically, customer requirement specifications are provided in semi-structured formats such as PDF or DOC files, which must be interpreted, structured, and categorized before they can be used effectively in development workflows or integrated into tools like IBM DOORS. 

Despite its importance, the current approach to handling requirement documents remains largely manual and error-prone. Extraction of requirements involves interpreting diverse document layouts, identifying section structures, and mapping content into a predefined schema, all of which require significant human effort. Following extraction, classifying the requirements into categories such as functional, non-functional, informational, or structural headers is similarly labor intensive and time consuming. These manual steps introduce delays, reduce traceability, and impact the overall quality of the RE process.

To address these challenges, this work introduces ReXCL, a tool designed to automate both the extraction and classification of requirement texts. The goal is to reduce human effort, increase consistency and streamline the transformation of raw requirement documents into structured, actionable data formats suitable for downstream RE tools and processes.

\section{Tool Description}
The ReXCL tool, depicted in Figure~\ref{fig:overall_workflow}, consists of two primary modules: Extraction and Classification. We define each component as follows:

\subsection{Extraction Module}
When provided with an input document, the Extraction module parses the raw document to extract the \emph{section numbers} (e.g., 1.1, 1.1.3), the corresponding \emph{section headings} (e.g., Main Task, Use Case), and the associated \emph{section texts} for each section \emph{i}. The components are merged into a structured format to produce the final extraction output of the input document. 


We refer to \emph{\( \text{section title}_i \)} as a combination of section \emph{\( \text{number}_i \)} and \emph{\( \text{heading}_i \)} (e.g., 1.1 Main Task), \emph{m} denotes the number of section texts associated with a section \emph{\( \text{title}_i \)} and \emph{n} denotes the number of section \emph{titles} present in the document. 


The process begins with the transformation of the input documents into an intermediate markdown-based text representation.  To eliminate noise from headers and footers (containing page numbers, dates, etc.), a lightweight Random Forest classifier is trained on labeled sentences using two features: frequency of occurrence and positional information.  Subsequently, the titles of the sections and their corresponding texts are identified and grouped into structured tuples of the form $\left< number_i, heading_i, [text_{i_{1}}, text_{i_{2}} ... , text_{i_{m}}] \right >$. These tuples are then formatted into a tabular schema and subsequently forwarded to the classification module for further processing.


\subsection{Classification Module}

This module performs requirement type classification by assigning the section texts to one of four categories: Info, Header, Functional Requirement, or Non-Functional Requirement. It leverages adaptive fine-tuning of BERT~\cite{devlin2019bertpretrainingdeepbidirectional}, customized for the automotive domain. The process begins with domain adaptation, where a pretrained general-purpose BERT model is further refined using large-scale unlabeled automotive requirement texts. This stage employs Iterative Masked Language Modeling to improve the model’s ability to capture domain-specific terminology and syntactic structures. 

Following domain adaptation, the model is fine-tuned in a supervised setting using labeled requirement texts. Each text is associated with a specific category, and the model is trained using a multi-layer perceptron, applied to the [CLS] token of the transformer's output, which captures document-level semantic representation. This approach enables the model to distinguish between closely related classes, particularly functional and non-functional requirements, while maintaining strong generalization across varied automotive documents. 

\section{Evaluation }
To assess the effectiveness of the Extraction module, three domain experts independently reviewed 445 sentences to assess if they correctly fit the  $\left< number_i, heading_i, text_i \right >$. As shown in Table~\ref{tab:extr_eval}, the accuracy is 4.45/5. The header/footer classification module achieved an average accuracy of 96.05\%, validating the reliability of the lightweight Random Forest model used for noise removal. 


Empirical results for the Classification module are in Table~\ref{tab:class_eval}. We find substantial performance gains from adaptive fine-tuning. It significantly outperforms the baseline vanilla BERT model (which yielded F1-scores between 0.22 and 0.41). The improvements are particularly notable in under-represented categories such as Non-Functional Requirements.

\begin{table}[!thb]
\caption{Manual Evaluation of the Extraction module}
\label{tab:extr_eval}
\centering
\small
\begin{tabular}{|c|cc|}
\hline
\multirow{2}{*}{\textbf{Expert}} & \multicolumn{2}{c|}{\textbf{445 sentences}} \\ \cline{2-3} 
 & \multicolumn{1}{c|}{\textbf{\begin{tabular}[c]{@{}c@{}}Extraction\\ Accuracy (/5)\end{tabular}}} & \textbf{\begin{tabular}[c]{@{}c@{}}Header-Footer\\ Accuracy (\%)\end{tabular}} \\ \hline
E1 & \multicolumn{1}{c|}{4.44} & 96.05 \\ \hline
E2 & \multicolumn{1}{c|}{4.44} & 96.05 \\ \hline
E3 & \multicolumn{1}{c|}{4.48} & 96.05 \\ \hline
Average & \multicolumn{1}{c|}{4.45} & 96.05 \\ \hline
\end{tabular}
\end{table}

\begin{table}[!thb]
\caption{Empirical evaluation of the Classification module}
\label{tab:class_eval}
\centering
\small
\begin{tabular}{|c|cc|}
\hline
\multirow{2}{*}{\textbf{Class Label}} & \multicolumn{2}{c|}{\textbf{Classification FScore}} \\ \cline{2-3} 
 & \multicolumn{1}{c|}{\textbf{Vanilla BERT}} & \textbf{\begin{tabular}[c]{@{}c@{}}Adaptive Finetuned \\ BERT\end{tabular}} \\ \hline
Header & \multicolumn{1}{c|}{0.41} & \textbf{0.99} \\ \hline
Info & \multicolumn{1}{c|}{0.4} & \textbf{0.98} \\ \hline
Func\_Req & \multicolumn{1}{c|}{0.24} & \textbf{0.98} \\ \hline
Non\_Func\_Req & \multicolumn{1}{c|}{0.22} & \textbf{0.93} \\ \hline
\end{tabular}
\end{table}

\section{Discussion of Related Work}
Although prior research has addressed header detection in documents~\cite{heading}, to the best of our knowledge, no existing work has holistically automated the schematization of semi-structured requirement documents that is being performed in the Extraction module. 

Recent work has investigated the application of BERT to the classification of requirement texts through transfer learning~\cite{arora2021classification, ajagbe2022retraining}. The studies highlight that domain-specific adaptation of language models significantly improves classification performance on requirement texts. We adopt an approach similar to ~\cite{stollenwerk2022adaptive} where models are first further pretrained on unlabeled domain-specific corpora before being fine-tuned on task-specific labeled data. This two-phase training process improves the model's ability to capture domain-specific nuances and handle low-resource classification tasks effectively.

While related work has addressed header detection and section classification separately for requirement documents, a comprehensive pipeline that automates both the structural schematization and classification of semi-structured requirement documents remains largely unexplored. 

\section{Conclusion and Future Work}
ReXCL automates extraction and classification in requirements engineering, improving efficiency and accuracy. It processes raw documents and assigns requirement types using heuristics and predictive modeling. Evaluation results validate ReXCL’s effectiveness in real-world settings. Future work includes extending support for images, tables, and multi-modal documents like presentations and spreadsheets.

\bibliographystyle{IEEEtran} 
\bibliography{biblio.bib}

@misc{heading,
      title={A Supervised Learning Approach For Heading Detection}, 
      author={Sahib Singh Budhiraja and Vijay Mago},
      year={2018},
      eprint={1809.01477},
      archivePrefix={arXiv},
      primaryClass={cs.IR},
      url={https://arxiv.org/abs/1809.01477}, 
}

@article{ser1,
  title={The role of requirement engineering in software development life cycle},
  author={Chakraborty, Abhijit and Baowaly, Mrinal Kanti and Arefin, Ashraful and Bahar, Ali Newaz},
  journal={Journal of emerging trends in computing and information sciences},
  volume={3},
  number={5},
  year={2012}
}

@inproceedings{ajagbe2022retraining,
  title={Retraining a BERT model for transfer learning in requirements engineering: A preliminary study},
  author={Ajagbe, Muideen and Zhao, Liping},
  booktitle={IEEE 30th International Requirements Engineering Conference (RE)},
  pages={309--315},
  year={2022},
  organization={IEEE}
}

@inproceedings{devlin2019bertpretrainingdeepbidirectional,
       title={{Bert: Pre-training of deep bidirectional transformers for language understanding}},
  author={Devlin, Jacob and Chang, Ming-Wei and Lee, Kenton and Toutanova, Kristina},
  booktitle={{2019 Conference of the North American Chapter of the Association for Computational Linguistics}},
  pages={4171--4186},
  year={2019}
}

@article{arora2021classification,
  title={Classification of software requirements using BERT-based deep learning},
  author={Arora, Anjali and Sikka, G.},
  journal={Computer Standards \& Interfaces},
  year={2021}
}

@article{stollenwerk2022adaptive,
  title={Adaptive Fine-Tuning of Transformer-Based Language Models for Named Entity Recognition},
  author={Stollenwerk, Felix},
  journal={arXiv preprint arXiv:2202.02617},
  year={2022}
}

\renewcommand{\appendixname}{Annex}
\appendix
This annex provides an overview of the interface and functionalities of the ReXCL tool along with more technical details of its constituent modules. A video of the deployed tool in action can be found at \href{https://youtu.be/G-GvWyhJaec}{ReXCL Demo}

\subsection{Deployment}
The frontend of the ReXCL tool has been developed using Angular, the backend is python FastAPI and the database used to store the documents and the intermediate results is MongoDB. The extraction module works comfortably in a CPU while the classification module requires a GPU. The tool therefore is currently deployed in a GPU server of 6GB RAM. 

\begin{figure}[!thb]
\caption{ReXCL: Dashboard}
\label{fig:dashboard}
\centering
\includegraphics[width=\linewidth]{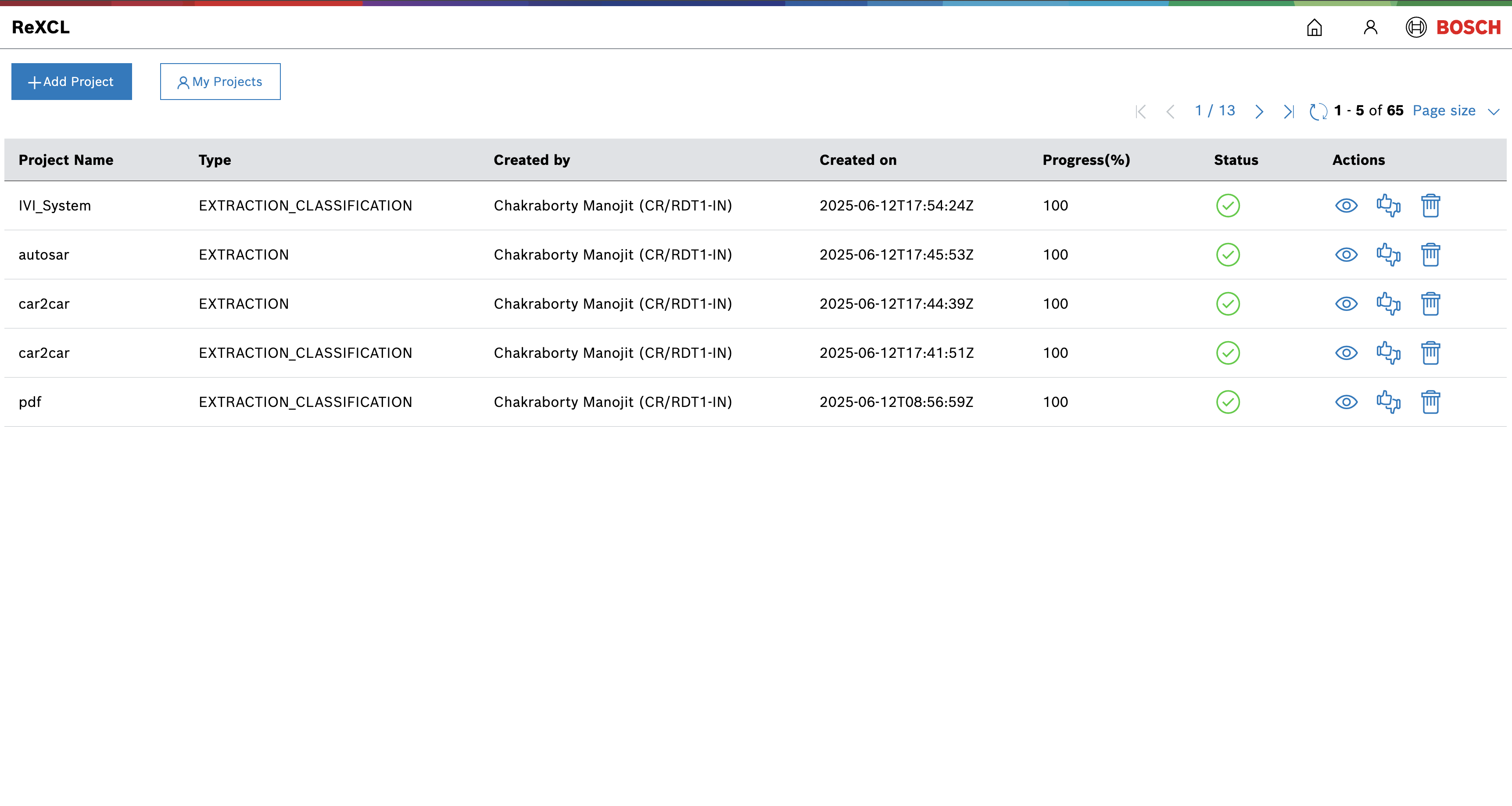}
\end{figure}

\begin{figure}[!thb]
\caption{ReXCL: View result from the Extraction module}
\label{fig:extraction_view}
\centering
\includegraphics[width=\linewidth]{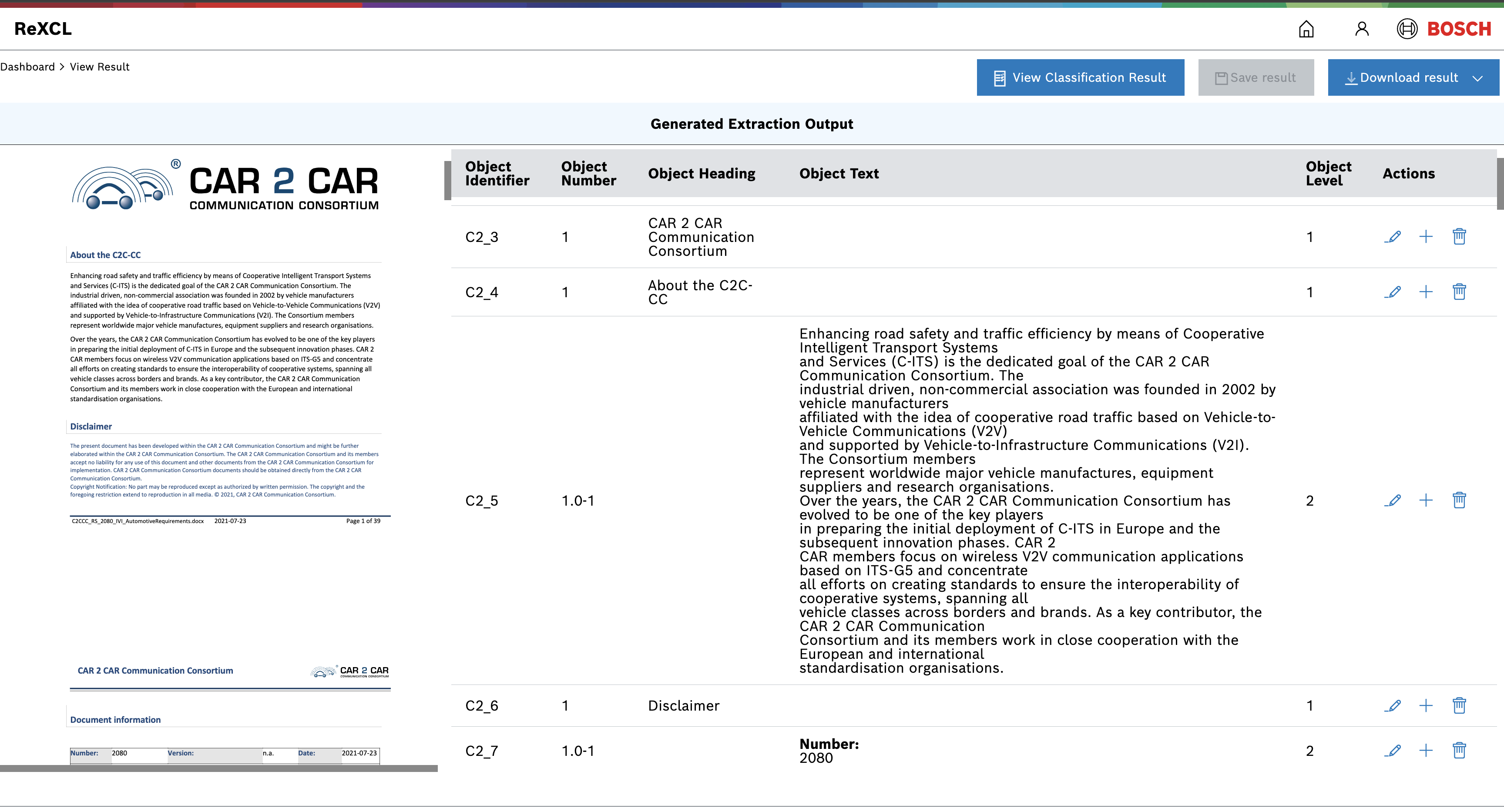}
\end{figure}

\begin{figure}[!thb]
\caption{ReXCL: Modify results from Extraction module in presence of inconsistencies}
\label{fig:extraction_edit}
\centering
\includegraphics[width=\linewidth]{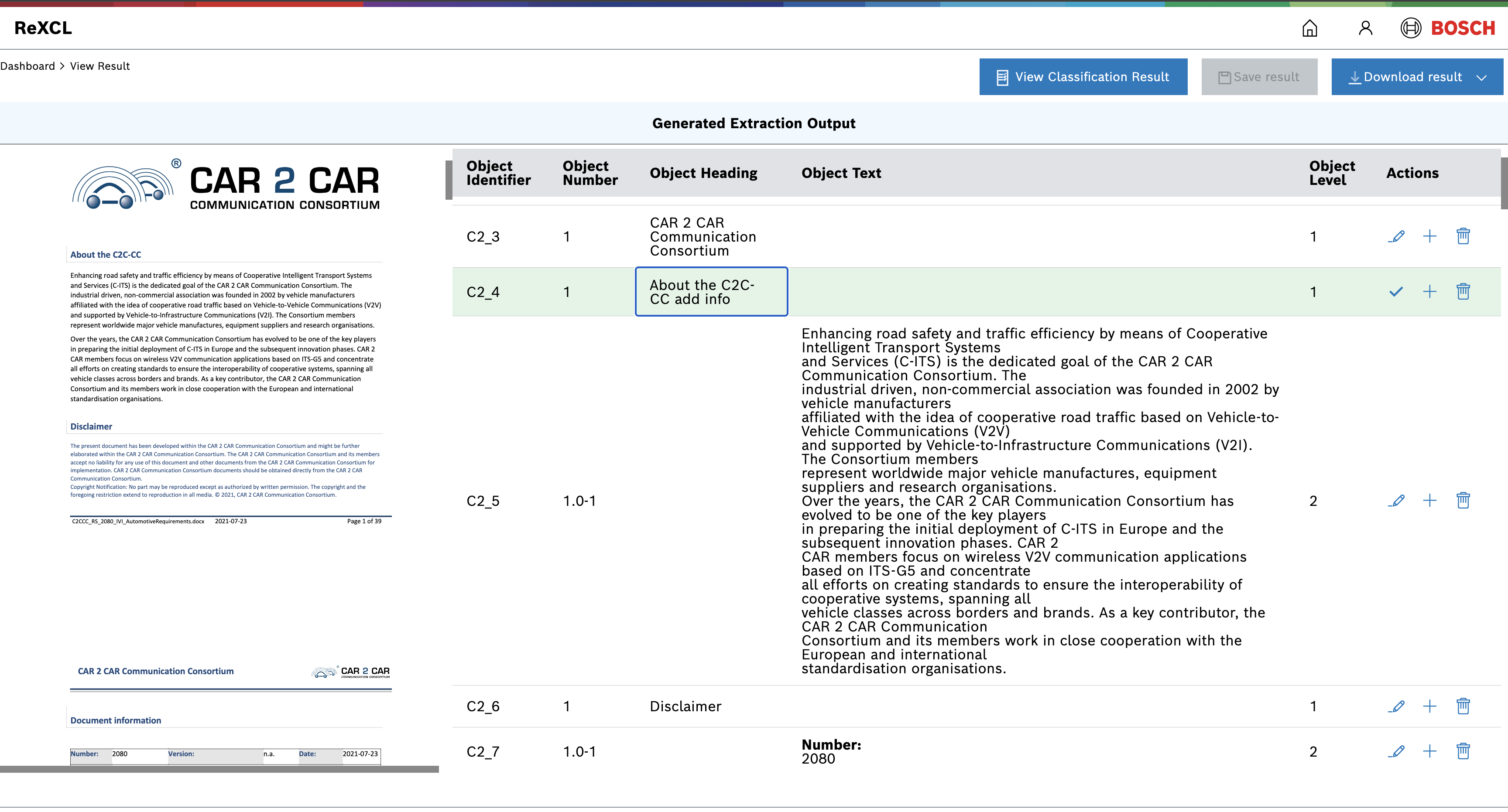}
\end{figure}

Figure~\ref{fig:dashboard} shows the dashboard of the ReXCL tool. One can view the results of the  previously uploaded requirement documents for extraction and classification. To start with the tool, one can either perform Extraction or Classification alone, or may execute the full pipeline. To perform extraction, we upload a file. The extraction module runs in the background and the result can be viewed, as shown in Figure~\ref{fig:extraction_view}. The right hand pane shows the structured information containing Object Identifier, Object Number (derived from the section numbers), Object Heading (derived from the section headings), Object Text (the corresponding text associated in the section) and Object Level (derived from the object number column using heuristics). The header and footers do not appear in the output. If the user finds some section information (e.g. numbers, headings, text, level) to be erroneously extracted, the user has the option to modify and correct and the result using the "Action" tab as displayed in Figure ~\ref{fig:extraction_edit}. The ``Download'' button on the top-right enables the user to download the result in different formats like csv, excel, yaml and json. This output can be used independently by the user or he/she can proceed to the "Classification" module, which we describe next.

\begin{figure}[!thb]
\caption{ReXCL: View result from the Classification module}
\label{fig:classification_view}
\centering
\includegraphics[width=\linewidth]{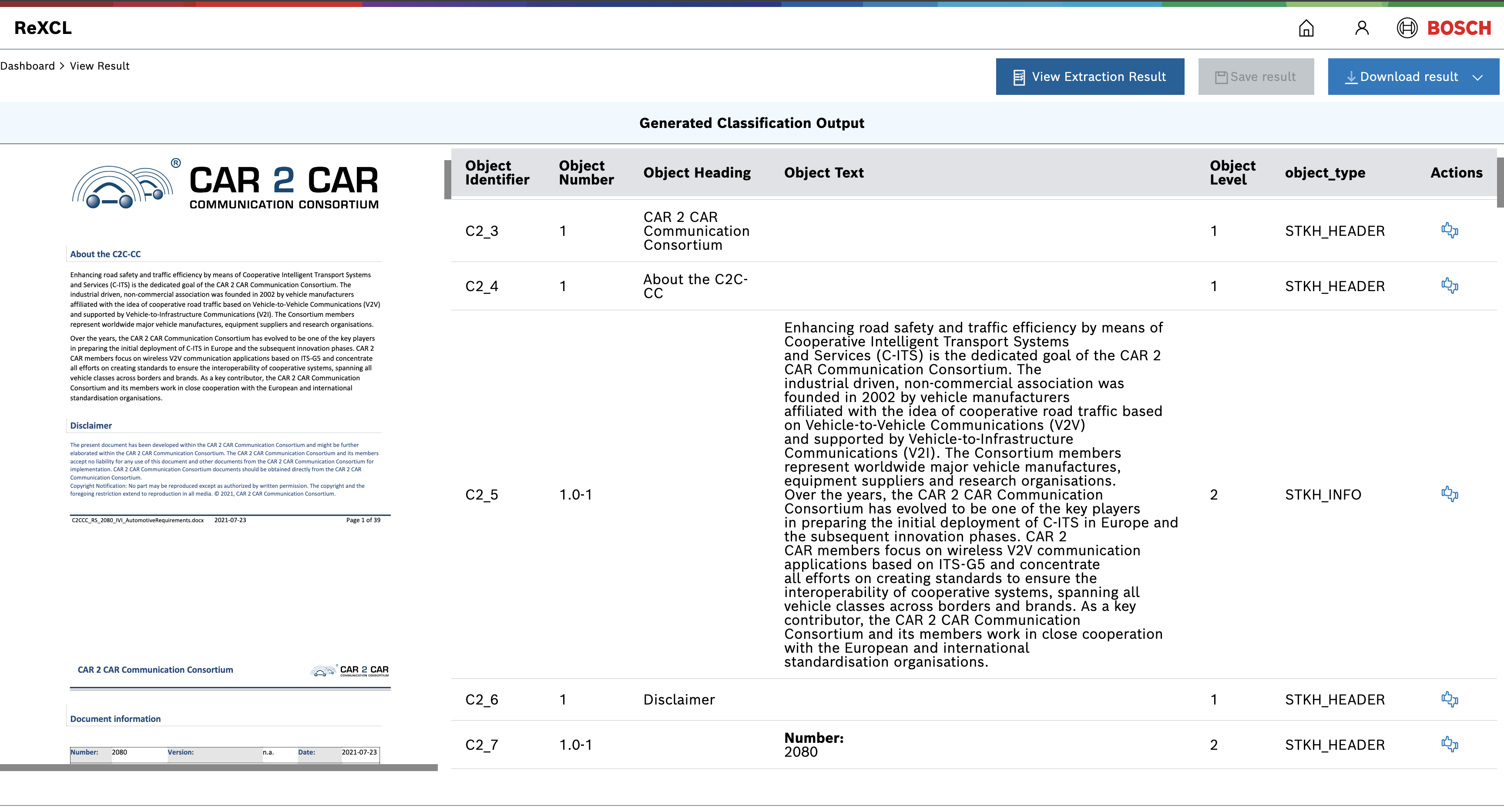}
\end{figure}

\begin{figure}[!thb]
\caption{ReXCL: Modify results from Classification module in presence of inconsistencies}
\label{fig:classification_edit}
\centering
\includegraphics[width=\linewidth]{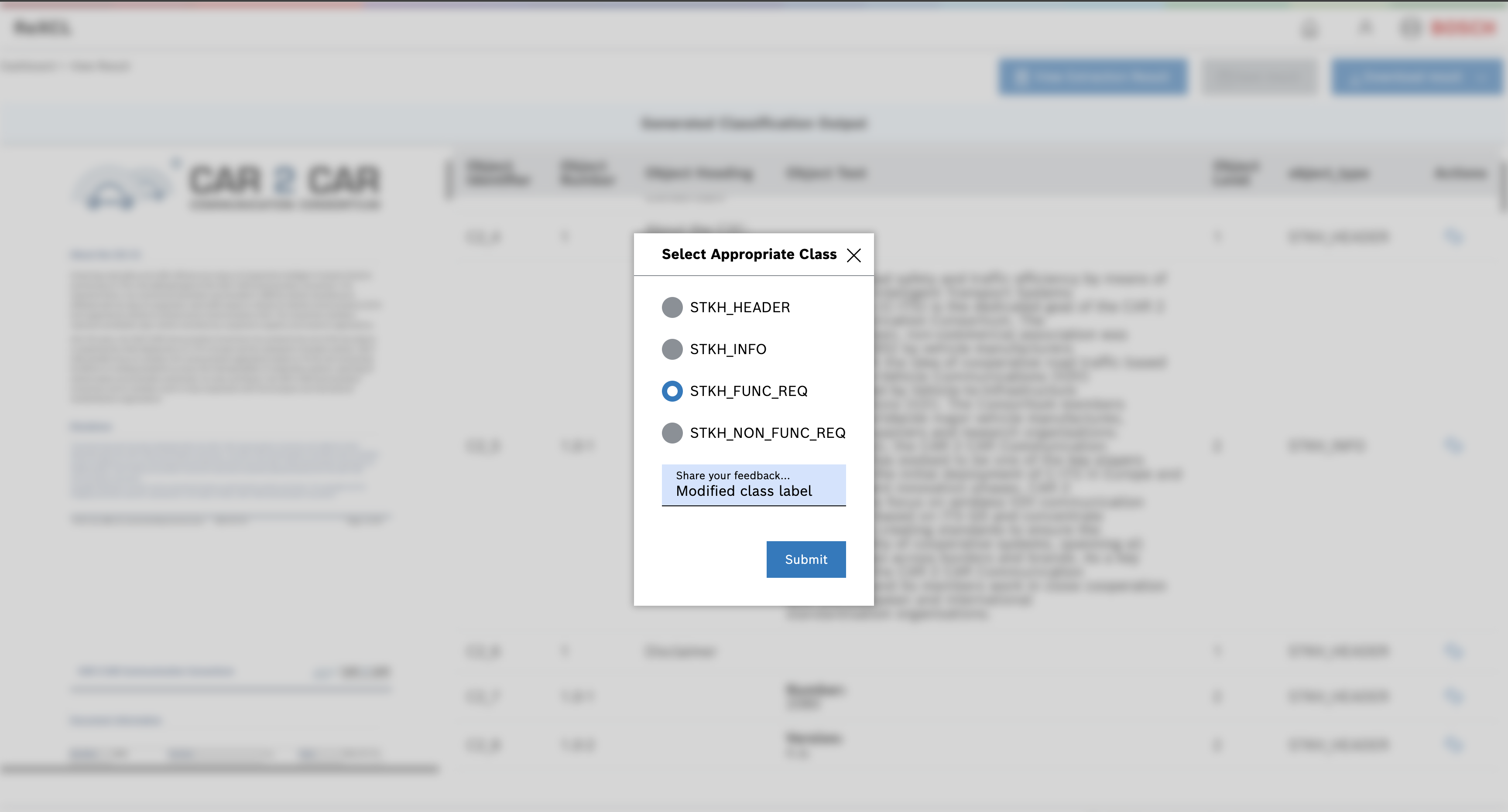}
\end{figure}

In the Classification module, each Object text is labeled using an Object Type which is one of the labels mentioned in Section 2.2. This is shown in Figure~\ref{fig:classification_view}. The feedback is captured through ``Action'' where the user can mention if the classifier output is correct or incorrect. In case the label is incorrect, the tool prompts the user to provide the correct label which is then saved, as shown in Figure~\ref{fig:classification_edit}. Similar to the extraction module, the results for classification can be downloaded in a structured format, which now contains the "Object Type" information.

The final output of the tool can be exported to RE tools like IBM DOORS or Jira in CSV, Excel, or JSON formats.

\subsection{Details of the Extraction Module}
The aim of the Extraction module is to convert unstructured requirement documents (pdf, doc, txt) into a structured format with Section number, Section heading and associated section texts. The key steps in the extraction process are :

\noindent
\textbf{1. Intermediate Text Representation} : Using regular text conversion packages like PyMuPDF converts all the contents into text. inducing loss of document structural information, styles and flattened tables. We use markdown text conversion packages like PyMuPDF4LLM which converts documents into markdown format. This preserves document structure by marking section titles with special tokens (e.g., \#), retaining table formatting and image captions.

\begin{figure}[]
\caption{The extraction module workflow; the input is a raw document, and the output is a final structured output containing section number, section heading and section text. The components used are intermediate text representation, header-footer removal, section information extraction and final output generation.}
\label{fig:extraction_workflow}
\centering
\includegraphics[width=\linewidth]{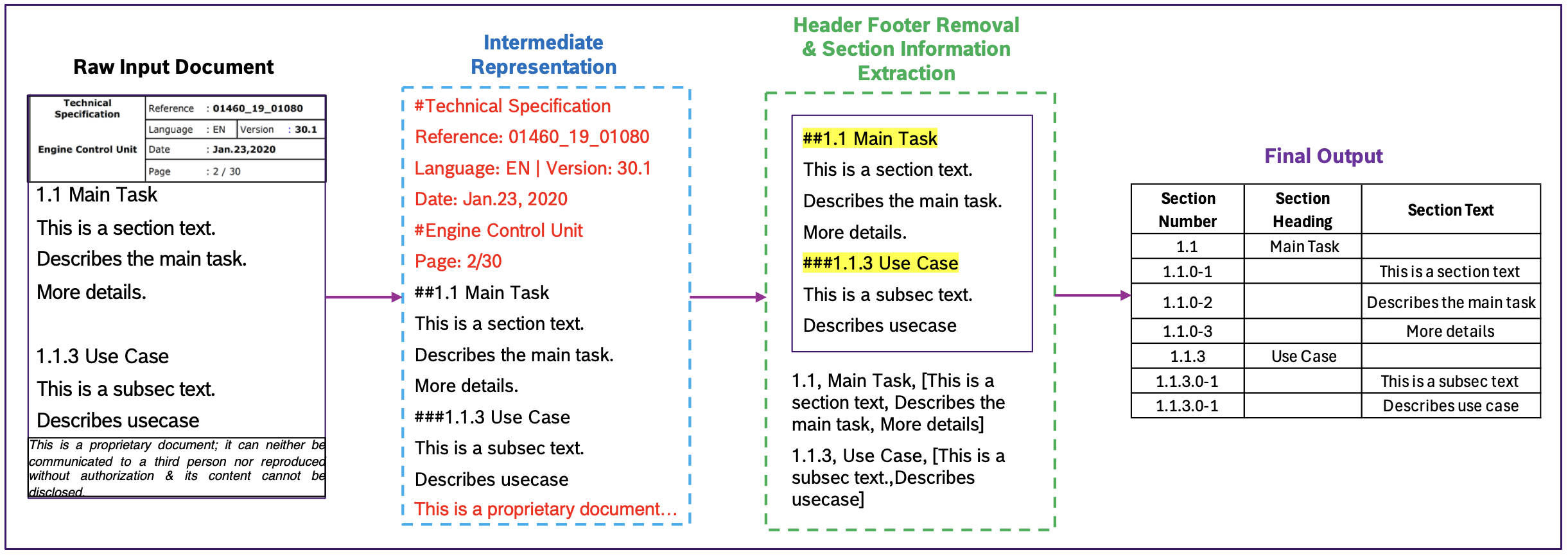}
\end{figure}

\noindent
\textbf{2. Header-Footer Removal}: In the intermediate text representation, headers and footers appear as sentences (highlighted in red in Fig.\ref{fig:extraction_workflow}). We approach this as a binary classification problem, labeling each sentence as either header-footer or requirement text. A lightweight Random Forest classifier is developed using two features: $frequency$ and $position$. We hypothesize that header and footer texts are redundant, occurring multiple times, and have fixed positions--headers at the top and footers at the bottom of pages. We annotate a sample of three documents with $3,773$ sentences into the classes \textit{Header-Footer} or \textit{Req.Text}. This trained model is then employed to detect and remove headers and footers.  

\noindent
\textbf{3. Section Information Extraction}: 
Given a raw customer requirement, we generate its intermediate text representation, eliminate header-footers to obtain the final requirement text, which is then parsed to extract section titles (section number and section heading) and section text (highlighted in yellow in Fig.~\ref{fig:extraction_workflow}). The final output is a list of tuples $\left< number_i, heading_i, [text_{i_{1}}, text_{i_{2}} ... , text_{i_{m}}] \right >$

\noindent
\textbf{4. Final Output Generation}: This submodule structures the extracted tuples into a tabular format, ready for classification and export to tools like IBM DOORS.


%

\subsection{Details of the Classification Module}
This module classifies requirement texts into specific categories: Info, Header, Functional Requirement, and Non-Functional Requirement. It utilizes advanced adaptive fine-tuning methods on transformer-based language models, such as BERT~\cite{devlin2019bertpretrainingdeepbidirectional}, tailored specifically for the automotive domain to handle its unique challenges.

The process begins by refining general-purpose transformer models using extensive unlabeled automotive-specific datasets. This adaptive fine-tuning step aligns the model more closely with the unique distribution of automotive requirement documents. The preprocessing phase is crucial here; it involves converting text to lowercase, removing punctuation except for essential structural markers like underscores, and preserving key words such as negations (e.g., not) and modal verbs (e.g., should). This ensures important semantic details remain intact. Iterative Masked Language Modeling further refines the models by repeatedly masking random tokens within the unlabeled corpus, enhancing the model’s understanding of domain-specific vocabulary and sentence structures.

After this initial adaptation, the model undergoes supervised fine-tuning using labeled requirement datasets. In this stage, requirement texts are explicitly associated with their designated categories. The model employs a classification head - specifically, a multi-layer perceptron, that processes the transformer's [CLS] token output, which summarizes the semantic content of the text.
This structured method effectively discriminates between closely related categories, such as Functional and Non-Functional Requirements, and ensures robust generalization across diverse automotive documentation. Overall, this approach addresses the complexities inherent in accurately classifying automotive requirement types.

\begin{figure}[thb]
  \centering
  \caption{Requirement Classification using Adaptive Finetuning. Input is requirement documents with/without class labels. }
  \includegraphics[width=\linewidth]{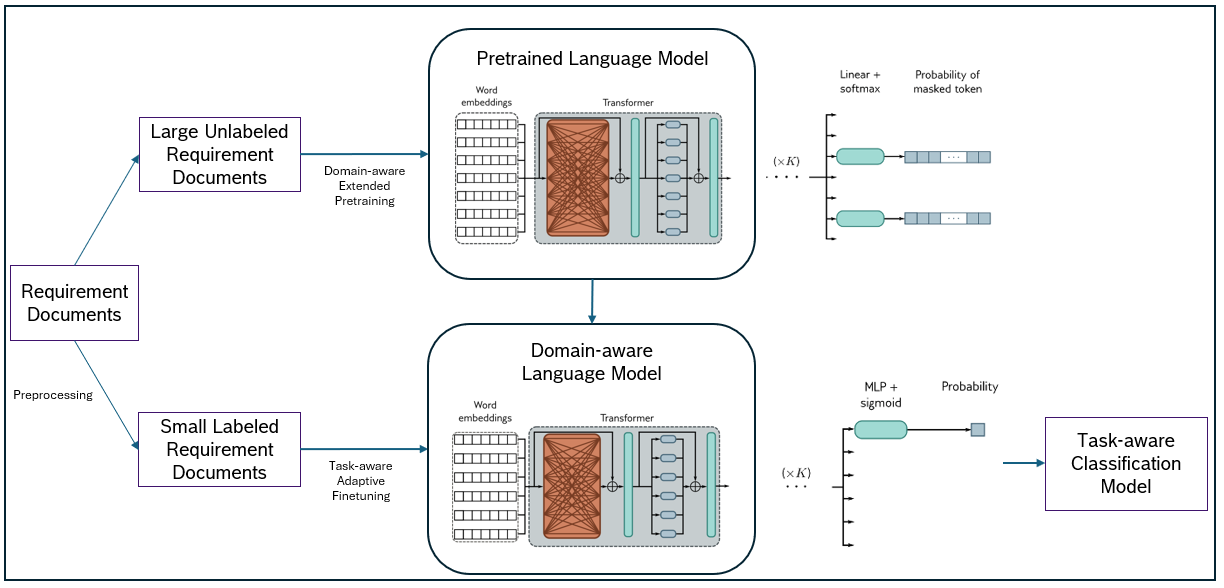}
  \label{fig:classification}

\end{figure}

\end{document}